\documentclass{article}

\usepackage[preprint]{neurips_2026}
\usepackage[utf8]{inputenc}
\usepackage[T1]{fontenc}
\usepackage{amsmath,amssymb}
\usepackage{booktabs}
\usepackage{tabularx}
\usepackage{array}
\usepackage{graphicx}
\usepackage{xcolor}
\usepackage{microtype}
\usepackage{enumitem}
\usepackage{float}
\usepackage{hyperref}
\usepackage[capitalize,noabbrev]{cleveref}
\usepackage{tikz}
\usetikzlibrary{arrows.meta,positioning,fit}

\hypersetup{
  colorlinks=true,
  linkcolor=blue!60!black,
  citecolor=blue!60!black,
  urlcolor=blue!60!black,
  pdftitle={Julia for CFD: A Critical Survey of Ecosystem, Performance, and Composability},
  pdfauthor={Tianbai Xiao},
  pdfkeywords={computational fluid dynamics, Julia, scientific computing, high-performance computing, GPU computing, automatic differentiation, scientific machine learning, differentiable programming}
}

\newcommand{\pkg}[1]{\texttt{#1}}

\title{Julia for CFD: A Critical Survey of Ecosystem, Performance, and Composability}

\author{%
  Tianbai Xiao \\
  Institute of Mechanics, Chinese Academy of Sciences \\
  Beijing, China \\
  \texttt{txiao@imech.ac.cn}
}

\begin{document}

\maketitle

\begin{abstract}
Modern computational fluid dynamics increasingly embeds forward simulation within broader workflows for design, inference, and data-driven modeling.
This evolution poses an architectural challenge: physical models, high-performance kernels, accelerator implementations, and data-driven tools often reside in separate languages or independently engineered software layers.
Julia offers a contrasting model in which high-level scientific abstractions can be specialized for performance and composed with differentiation and learning within a common language and compiler ecosystem.

This critical survey asks \emph{where a single-language, composable software model reduces architectural impedance in CFD, and where it does not}. It examines representative open-source projects and the broader Julia infrastructure that supports them.
Instead of relying on language-level microbenchmarks, we synthesize published application-level evidence on performance, scaling, accelerator portability, differentiability, and software composition. The literature demonstrates credible Julia-native CFD on large distributed CPU systems and multi-GPU platforms, alongside emerging differentiable workflows.

Comparisons with C++ performance-portability frameworks, finite-element domain-specific languages (DSLs), and JAX-based differentiable CFD show that GPU portability, AD, and composability are not unique to Julia. Julia's distinction lies in bringing these capabilities into a shared programming and specialization model. The evidence is nonetheless mixed: Julia has moved beyond proof of concept in several CFD regimes, but its ecosystem still lacks the breadth, industrial tooling, and deployment experience of established C/C++/Fortran environments. Its strongest role may therefore be as a platform for exploring CFD architectures that integrate simulation with downstream analysis, rather than as a universal replacement language.
\end{abstract}

\noindent\textbf{Keywords:} Julia; computational fluid dynamics; scientific computing; differentiable programming; scientific machine learning

\section{Introduction}
\label{sec:introduction}

Computational fluid dynamics (CFD) has always been shaped by both numerical analysis and computing technology. Although its governing equations can be written compactly, a practical CFD solver is a complex software system that must connect physical models and discretizations to solution algorithms, parallel execution, and data management. For several decades, high-performance implementations have relied primarily on Fortran, C, and C++, supported by mature compilers, parallel toolchains, and numerical libraries. MATLAB and Python often complement the core solver by handling setup, analysis, and workflow orchestration. This division is effective, but it can produce a layered stack in which performance-critical simulation and higher-level analysis evolve under different programming models. The architecture connecting these layers can influence which numerical ideas are practical almost as strongly as their formal accuracy or stability properties.

The cost of this separation grows as CFD moves beyond isolated forward simulations toward design, inference, and data-driven modeling. The solver then becomes one component in a larger computational process, with states and derivatives repeatedly passing between algorithms; a language boundary that is negligible for a single offline run may become intrusive inside an optimization loop or a differentiable simulation. Julia was designed in part to challenge this separation between high-level scientific programming and performance-oriented implementation. It combines an interactive language with compilation to native code, specialization, multiple dispatch, and generic programming \citep{Bezanson2017Julia}. Algorithms can be written against abstract interfaces while the compiler specializes concrete combinations of models, data types, and execution backends. When those interfaces are designed carefully, one implementation can often be reused across numerical representations and hardware targets. Assessing this model therefore requires evidence on kernel performance and on how complete workflows are composed, transformed, and maintained.

CFD provides a demanding test of this model. Credible software must handle both regular and irregular data structures, scale across distributed and heterogeneous systems, and remain extensible as numerical methods and physical models change. Source-level elegance is of little value if it conceals prohibitive memory traffic or communication costs; conversely, kernel performance alone does not ensure that a solver can be differentiated, coupled to learned models, or adapted to new hardware. There is now enough Julia-native fluid software to examine these tradeoffs at application scale. Trixi.jl, Oceananigans.jl, and WaterLily.jl represent distinct numerical and physical settings \citep{Ranocha2022Trixi,Ramadhan2020Oceananigans,Weymouth2025WaterLily}, while other projects extend to engineering, atmospheric, kinetic, and multiphase simulation. Published studies report large distributed CPU calculations and multi-GPU applications, and several solvers expose differentiable or backend-portable designs \citep{Candelaresi2026MassivelyParallel,Font2026WaterLilyScaling,Silvestri2023OceananigansScaling,Agdestein2026INS,LCS2026}. This evidence shows that the ecosystem has moved beyond small demonstrations, although maturity remains uneven across CFD.

Meaningful assessment also requires comparison with architectures already used elsewhere. Established Fortran and C++ codes derive much of their strength from mature domain libraries, compiler toolchains, and long-validated parallel implementations. Newer frameworks layer higher-level abstractions over this foundation in different ways. AMReX provides reusable adaptive-mesh and performance-portable infrastructure in C++ \citep{Zhang2019AMReX}; Firedrake couples high-level finite-element problem descriptions to generated kernels and established solver libraries \citep{Rathgeber2016Firedrake}; and JAX-based packages build differentiable, accelerator-oriented CFD on array programming and automatic differentiation (AD) \citep{Kochkov2021JAXCFD,Bezgin2025JAXFluids2,FAN2026118455}. These systems provide three relevant baselines: reusable C++ infrastructure for adaptive and heterogeneous computation, symbolic generation of finite-element kernels, and array-based differentiable CFD. Julia's distinction is narrower. Its generic types, multiple dispatch, and compiler specialization are available directly to application and library code, so numerical components and program transformations can, in principle, share the same interfaces. The following sections examine whether this arrangement actually reduces glue code and duplicated representations once performance portability, automatic differentiation, and external libraries are required.

The central thesis of this survey concerns software architecture. We use \emph{architectural impedance} to describe the engineering friction introduced when related numerical tasks are separated by incompatible languages, representations, or transformation systems. Julia's distinctive experiment is whether physical models, efficient implementations, hardware portability, and algorithmic differentiation can participate in a common type and specialization framework. Such composition is not automatic: complexity may reappear as compilation latency, package incompatibility, or incomplete backend and AD support. The relevant measure is whether this approach lowers the overall cost of extending CFD software across new models, hardware, and modes of analysis. To assess that proposition, this survey examines Julia's language mechanisms, maps representative fluid-dynamics software, and evaluates published evidence on performance and composability in comparison with adjacent approaches. The emphasis is therefore on a practical architectural question: when does a shared software environment for simulation, differentiation, optimization, and data analysis offer a real advantage over extending an established CFD solver?

\section{Scope and survey methodology}
\label{sec:scope}

This article serves as a \emph{critical narrative survey}. It considers representative, actively developed, open-source software in which Julia is responsible for a substantive part of the CFD simulation. The scope includes dedicated CFD solvers, broader PDE frameworks, particle and kinetic methods, and shared numerical or HPC infrastructure when it materially affects solver architecture. Educational examples, inactive projects, and thin wrappers around solvers implemented elsewhere are omitted unless they illuminate a distinct design pattern.

The evidence base was screened through August 2026 and combines archival publications, official documentation, and public repositories. Priority was given to projects with documented fluid-dynamics capability, validation or benchmark evidence, an open development history, and relevance to performance, heterogeneous execution, differentiation, or software composition. Because documentation and maturity vary, repository claims are used to describe features, whereas quantitative performance claims are drawn from citable studies whenever possible.

Projects are examined along three axes. The first records their physical and numerical setting; the second records demonstrated capabilities such as distributed or accelerator execution and differentiation; the third examines where software boundaries occur and which components can be exchanged without duplicating solver logic. Performance studies use different equations, hardware, precision, and scaling protocols, so their results are treated as \emph{capability evidence}, not as a normalized ranking. Comparisons with AMReX, Firedrake, and JAX-based CFD provide external reference points for judging which properties arise from Julia and which reflect broader trends in scientific software.

\section{Why Julia is relevant to CFD software design}
\label{sec:why_julia}

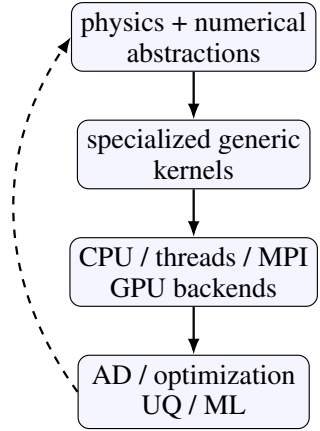
\begin{figure}[t]
\centering
\resizebox{0.98\textwidth}{!}{%
\begin{tikzpicture}[
  node distance=6mm and 12mm,
  box/.style={draw, rounded corners, align=center, minimum width=2.9cm, minimum height=8mm, fill=blue!4},
  arrow/.style={-{Latex[length=2mm]}, thick},
  title/.style={font=\bfseries}
]
\node[title] (t1) {Conventional multi-language CFD stack};
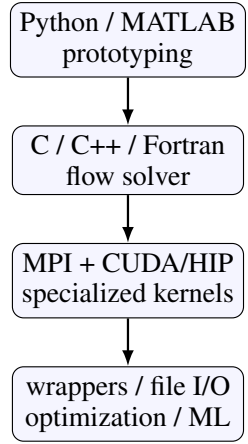
\node[box, below=5mm of t1] (proto) {Python / MATLAB\\prototyping};
\node[box, below=of proto] (solver) {C / C++ / Fortran\\flow solver};
\node[box, below=of solver] (accel) {MPI + CUDA/HIP\\specialized kernels};
\node[box, below=of accel] (ext) {wrappers / file I/O\\optimization / ML};
\draw[arrow] (proto) -- (solver);
\draw[arrow] (solver) -- (accel);
\draw[arrow] (accel) -- (ext);

\node[title, right=20mm of t1] (t2) {Julia-native composable stack};
\node[box, below=5mm of t2] (physics) {physics + numerical\\abstractions};
\node[box, below=of physics] (kernels) {specialized generic\\kernels};
\node[box, below=of kernels] (backend) {CPU / threads / MPI\\GPU backends};
\node[box, below=of backend] (learn) {AD / optimization\\UQ / ML};
\draw[arrow] (physics) -- (kernels);
\draw[arrow] (kernels) -- (backend);
\draw[arrow] (backend) -- (learn);
\draw[arrow, dashed] (learn.west) to[bend left=35] (physics.west);
\end{tikzpicture}%
}
\caption{A schematic comparison of a conventional multi-language CFD workflow and a Julia-native composable workflow. These architectural tendencies can coexist, and mature applications in either model may call external libraries or combine languages.}
\label{fig:stack}
\end{figure}

Kernel performance is only one part of Julia's relevance to CFD software design. Modern C++, Fortran, accelerator libraries, and domain-specific systems can all produce efficient kernels.
The broader question is how a programming model relates physical models, numerical methods, data representations, hardware backends, and transformations such as differentiation. Julia is most relevant when these relationships can be expressed through ordinary language mechanisms and their concrete combinations specialized within a shared programming environment.
The workflow contrast is sketched in \cref{fig:stack}, and the Julia mechanisms discussed below are summarized in \cref{tab:julia_mechanisms}.

\subsection{From two-language workflows to composable numerical software}

A conventional workflow may prototype a model in one language, implement production kernels in another, and add GPU or inverse-problem support through further layers. This division is often rational, but it becomes costly when a changing state representation or numerical operator must be reflected across several interfaces. Julia's response is multiple dispatch combined with compiler specialization: behavior can be defined for interactions among equation, mesh, state, and backend types, while concrete combinations are compiled when used. This organization matches CFD, where an operation usually belongs to a combination of numerical and physical objects rather than to one object in isolation.

As an example, Trixi.jl illustrates the resulting form of reuse. Equations, fluxes, meshes, semidiscretizations, and analysis tools are separated by generic interfaces \citep{Ranocha2022Trixi}; TrixiLW.jl can therefore introduce a different time-space discretization while retaining curved-mesh and postprocessing infrastructure instead of maintaining a fork \citep{Babbar2024TrixiLW}. Generic numerical code can also admit alternative precisions, device arrays, or dual numbers when it avoids assumptions tied to a single scalar or storage type \citep{Revels2016ForwardDiff}. Variants can thus share a semantic interface even when each still requires implementation work.

This flexibility redistributes low-level work. Excessive specialization increases compilation time and code size, type instability can reintroduce runtime overhead, and a clean abstraction may still generate poor memory access. Effective Julia CFD therefore requires \emph{abstraction engineering}: interfaces must be broad enough to reuse and concrete enough to compile into predictable kernels. The benefit is realized only when a new model or backend can reuse verified numerical structure without sacrificing visibility into performance.

\begin{table}[t]
\centering
\caption{Julia mechanisms that are directly relevant to CFD software design. The final column records the performance or maintenance constraints associated with each mechanism.}
\label{tab:julia_mechanisms}
\small
\setlength{\tabcolsep}{4pt}
\renewcommand{\arraystretch}{1.12}
\begin{tabularx}{\textwidth}{>{\raggedright\arraybackslash}p{2.8cm} >{\raggedright\arraybackslash}p{3.4cm} X >{\raggedright\arraybackslash}p{3.6cm}}
\toprule
\textbf{Mechanism} & \textbf{CFD use pattern} & \textbf{Potential benefit} & \textbf{Main caveat} \\
\midrule
Multiple dispatch & Equations, fluxes, boundary conditions, meshes, closures & Separates interacting concepts without rigid inheritance; enables package extensions to reuse numerical components & Method ambiguities and interface conventions must be controlled as ecosystems grow \\
\addlinespace
Type specialization & Precision, state types, small tensors, AD scalars & One generic algorithm can generate concrete machine code for different numerical representations & Compilation latency, code size, and type instability can become significant \\
\addlinespace
Generic arrays and kernels & CPU, GPU, distributed or custom storage & Reduces duplicated backend-specific source and keeps numerical logic close to the solver & Source portability does not guarantee comparable performance on every architecture \\
\addlinespace
First-class functions and metaprogramming & Flux construction, callbacks, generated operators, model composition & Makes numerical algorithms and transformations easy to express as ordinary program components & Compiler-aware metaprogramming can become difficult to maintain \\
\addlinespace
AD-compatible numerical code & Sensitivity analysis, inverse problems, learned closures & Allows differentiation to reuse parts of the production solver instead of a separate tensor implementation & Mutation, iterative solvers, MPI, AMR, shocks, and long-time chaos still require specialized treatment \\
\bottomrule
\end{tabularx}
\end{table}

\subsection{Parallelism, accelerators, and compilation model}

Julia CFD uses MPI, accelerator runtimes, and tuned vendor libraries as part of its execution stack. MPI.jl exposes standard implementations \citep{Byrne2021MPI}, GPU packages compile Julia kernels to device targets \citep{Besard2019GPU}, and solvers can call optimized FFT or linear-algebra libraries where appropriate. The architectural opportunity is to represent execution choices through array and backend types, allowing solver logic to remain shared while dispatch selects portable or architecture-specific kernels. Structured solvers offer the clearest examples, whereas unstructured meshes and migrating particles generally require more backend specialization.

Shared HPC infrastructure determines how far this approach can extend. Portable primitives for reductions, scans, sorting, and related operations reduce the need for every solver to build device-specific utilities \citep{Nicusan2025AcceleratedKernels,Pilliat2026KernelForge}. Nevertheless, source portability is not performance portability: data layout, launch structure, communication, and memory behavior remain application responsibilities. A useful common interface may therefore select distinct implementations on different devices when their performance requirements differ.

Specialization also changes deployment costs. The first encounter with a method combination may include compilation, so time to first solution differs from steady-state throughput. This cost is often amortized in a long run but can dominate short jobs or large ensembles; at scale, code loading and synchronized compilation can also stress shared filesystems \citep{Candelaresi2026MassivelyParallel}. Precompiled or system images move this cost into a build phase, trading interactive flexibility for a more fixed deployment artifact. Compilation must therefore be included in the workflow cost.

\subsection{Differentiation and solver composition}

AD tests composability more severely than ordinary library calls because it transforms the numerical program itself. Julia's AD ecosystem includes forward-mode dual numbers, reverse-mode systems based on source or compiler-IR transformation and recorded execution tapes, symbolic differentiation, and LLVM-level differentiation through Enzyme. DifferentiationInterface and ADTypes expose common derivative operators and allow a library to accept a backend rather than commit to one implementation \citep{Revels2016ForwardDiff,Moses2020Enzyme,Dalle2025DifferentiationInterface}. Custom derivative rules provide another layer: numerically significant operations such as linear solves or constitutive models can supply their own pushforwards or pullbacks instead of being differentiated instruction by instruction.

This variety matters in CFD because the appropriate method depends on both the requested derivative and the structure of the solver. Forward mode is effective for a small number of active parameters and local Jacobian-vector products, whereas reverse mode is attractive for scalar objectives with many parameters. Mutation, sparse operators, external library calls, device kernels, and value-dependent control flow affect the available backends differently, so a common interface does not imply uniform compatibility. At solver level, time integration and nonlinear solution are often differentiated through forward sensitivities, implicit rules, or continuous and discrete adjoints rather than by recording every iteration. Long trajectories introduce further choices about checkpointing, recomputation, and gradient stability. Julia's relevant capability is the possibility of selecting local derivative implementations and solver-level sensitivity algorithms separately within the same application.

These solver-level choices are exposed through the wider SciML ecosystem. DifferentialEquations.jl provides common interfaces for time integration \citep{Rackauckas2017DifferentialEquations}, while NonlinearSolve.jl combines nonlinear algorithms with linear solvers, sparsity information, and derivative backends \citep{Pal2024NonlinearSolve}. A CFD package can therefore delegate general time-integration and nonlinear-solution algorithms to these libraries by exposing its spatial semi-discretization and problem structure. This separation also supports scientific machine learning: a learned component can be embedded and trained within a mechanistic solve \citep{Rackauckas2020UDE}. In practice, the component must run on the target device, admit the required derivatives, and remain consistent with the discretization and solver. Julia reduces the interface work needed to assemble such a calculation in one program; conservation, stability, and identifiability remain properties of the model and numerical method.

\section{Mapping Julia CFD ecosystem}
\label{sec:ecosystem}

\begin{figure}[H]
\centering
\resizebox{0.98\textwidth}{!}{%
\begin{tikzpicture}[
  fam/.style={draw, rounded corners=2pt, align=center, text width=3.0cm, minimum height=1.9cm, fill=blue!4, font=\footnotesize},
  hub/.style={draw, very thick, rounded corners=4pt, align=center, text width=4.8cm, minimum height=1.5cm, fill=orange!12, font=\bfseries\small},
  infra/.style={draw, rounded corners=2pt, align=center, text width=4.35cm, minimum height=1.25cm, fill=green!5, font=\footnotesize},
  arr/.style={-{Latex[length=2mm]}, thick},
  darr/.style={-{Latex[length=2mm]}, thick, dashed}
]
\node[fam] (hyp) at (-6.8,3.1) {\textbf{High-order / hyperbolic}\\\pkg{Trixi.jl}};
\node[fam] (inc) at (-3.4,3.1) {\textbf{Structured incompressible}\\\pkg{WaterLily.jl}\\\resizebox{2.75cm}{\height}{\pkg{IncompressibleNavierStokes.jl}}};
\node[fam] (eng) at (0,3.1) {\textbf{Engineering / FEM}\\\pkg{XCALibre.jl}; \pkg{Gridap.jl}\\\resizebox{2.75cm}{\height}{\pkg{SegregatedVMSSolver.jl}}};
\node[fam] (geo) at (3.4,3.1) {\textbf{Geophysical / atmosphere}\\\pkg{Oceananigans.jl}\\\pkg{SpeedyWeather.jl}, \pkg{ClimaAtmos.jl}};
\node[fam] (kin) at (6.8,3.1) {\textbf{Kinetic / particle / multiphase}\\\pkg{Kinetic.jl}, \pkg{TrixiParticles.jl}\\\pkg{LCS.jl}};

\node[hub] (hub) at (0,0.45) {Julia-native CFD\\shared language, compiler,\\types, and dispatch};
\draw[arr] (hyp.south) -- (hub.north west);
\draw[arr] (inc.south) -- (hub.north west);
\draw[arr] (eng.south) -- (hub.north);
\draw[arr] (geo.south) -- (hub.north east);
\draw[arr] (kin.south) -- (hub.north east);

\node[infra] (exec) at (-5.0,-2.45) {\textbf{Execution}\\\pkg{MPI.jl}\\ \pkg{CUDA.jl} / \pkg{AMDGPU.jl} / \pkg{oneAPI.jl} / \pkg{Metal.jl}\\\pkg{KernelAbstractions.jl}};
\node[infra] (num) at (0,-2.45) {\textbf{Numerical infrastructure}\\\pkg{DifferentialEquations.jl} \\ \pkg{LinearSolve.jl} \\ \pkg{NonlinearSolve.jl} \\ \pkg{SparseArrays.jl}};
\node[infra] (diff) at (5.0,-2.45) {\textbf{Differentiation and learning}\\\pkg{ForwardDiff.jl} \\ \pkg{Zygote.jl} \\ \pkg{Enzyme.jl} \\ \pkg{Flux.jl} / \pkg{Lux.jl}};
\draw[darr] (exec.north) -- (hub.south west);
\draw[darr] (num.north) -- (hub.south);
\draw[darr] (diff.north) -- (hub.south east);
\end{tikzpicture}%
}
\caption{Representative Julia fluid-dynamics software grouped by numerical and application area. Solver families are organized around a shared Julia programming model, while the lower row shows infrastructure for parallel execution, numerical solution, differentiation, and learning. Solid and dashed arrows distinguish these two roles; they do not represent package dependencies.}
\label{fig:ecosystem_map}
\end{figure}
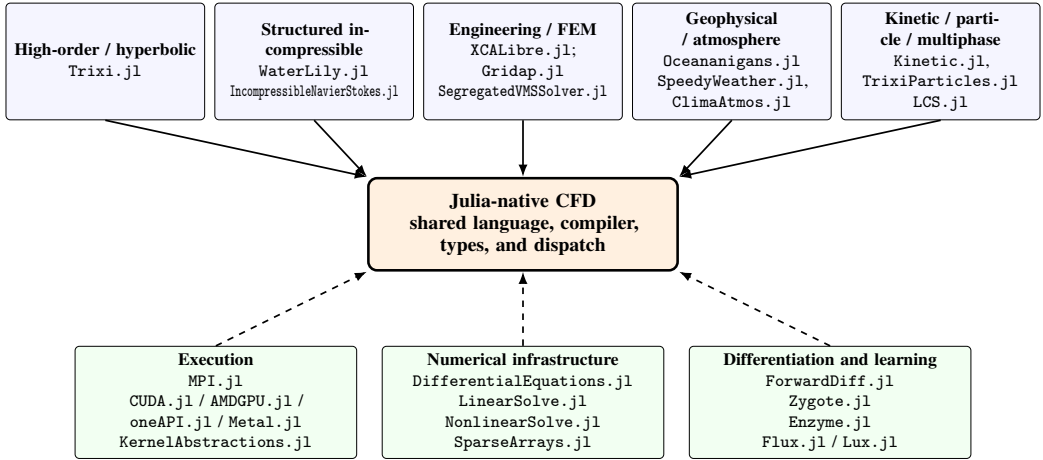

The projects summarized in \cref{fig:ecosystem_map,tab:ecosystem} solve different classes of problems using different discretizations and data structures. A direct comparison of their capabilities or performance would therefore be misleading: a regular-grid incompressible solver, a high-order framework, and a particle method are not intended for the same workloads. Accordingly, this section treats them as architectural case studies. Regular-grid packages illustrate portability and differentiation, high-order frameworks show how abstractions interact with specialization and adaptivity, and unstructured and particle codes expose the challenges of irregular memory access and communication. The figure and table summarize the scope of the ecosystem; the discussion that follows examines what each family reveals about the strengths and limits of Julia's programming model.

\begin{table}[p]
\centering
\caption{Representative Julia CFD projects grouped by numerical and application family. Entries summarize capabilities reported in publications or project documentation; availability may depend on configuration.}
\label{tab:ecosystem}
\scriptsize
\setlength{\tabcolsep}{3pt}
\renewcommand{\arraystretch}{1.12}
\begin{tabularx}{\textwidth}{>{\raggedright\arraybackslash}p{2.35cm} >{\raggedright\arraybackslash}p{2.35cm} >{\raggedright\arraybackslash}p{3.25cm} >{\raggedright\arraybackslash}p{2.35cm} X}
\toprule
\textbf{Family / project} & \textbf{Primary physics} & \textbf{Numerical / geometric emphasis} & \textbf{Execution} & \textbf{Architectural relevance} \\
\midrule
\pkg{Trixi.jl} & Hyperbolic systems; compressible CFD & High-order DG/DGSEM, entropy-stable methods, AMR & CPU, MPI; GPU work & Extensible equation/flux abstractions; high-order HPC \citep{Ranocha2022Trixi,Candelaresi2026MassivelyParallel} \\
\addlinespace
\pkg{WaterLily.jl} & Incompressible viscous flow around moving bodies & Cartesian immersed-boundary solver & CPU, GPU, MPI & Compact backend-agnostic and differentiable solver \citep{Weymouth2025WaterLily,Font2026WaterLilyScaling} \\
\addlinespace
\resizebox{2.2cm}{\height}{\pkg{IncompressibleNavierStokes.jl}} & Incompressible turbulence & Energy-conserving staggered Cartesian discretization & Threaded CPU, GPU & Matrix-free single-source kernels and discrete adjoints \citep{Agdestein2026INS} \\
\addlinespace
\pkg{XCALibre.jl} & Engineering incompressible / weakly compressible CFD; RANS/LES & Cell-centered finite volume on structured/unstructured meshes & Multithreaded CPU, multiple GPU backends & Tests Julia in irregular engineering-CFD workflows \citep{Medina2025XCALibre} \\
\addlinespace
\pkg{Gridap.jl} ecosystem & PDEs including Stokes/Navier--Stokes & Extensible finite elements; distributed and scalable solver layers & CPU, MPI via ecosystem & Generic weak forms and composable FEM infrastructure \citep{Badia2020Gridap,Manyer2024GridapSolvers} \\
\addlinespace
\resizebox{2.2cm}{\height}{\pkg{SegregatedVMSSolver.jl}} & Incompressible LES & Stabilized variational multiscale finite elements & Parallel Gridap ecosystem & Application-level CFD built compositionally on FEM infrastructure \citep{Brunelli2025SegregatedVMS} \\
\addlinespace
\pkg{Oceananigans.jl} & Oceanic / geophysical incompressible flow & Hydrostatic and nonhydrostatic finite volume & CPU, GPU, distributed & GPU-first geophysical CFD from idealized to global scale \citep{Ramadhan2020Oceananigans,Silvestri2023OceananigansScaling,Wagner2025Oceananigans} \\
\addlinespace
\pkg{SpeedyWeather.jl} & Atmospheric general circulation & Spectral / spherical-harmonic atmospheric modeling & CPU-oriented, interactive & Emphasizes interactivity and extensibility in global modeling \citep{Klower2024SpeedyWeather} \\
\addlinespace
\pkg{ClimaAtmos.jl} & Compressible atmospheric dynamics and parameterizations & Cubed-sphere/column grids; climate-model components & CPU, GPU, MPI & Physics-component composition, calibration, data assimilation and ML \citep{Sridhar2022ClimateMachine,Yatunin2026ClimaAtmos} \\
\addlinespace
\pkg{Kinetic.jl} & Kinetic, rarefied and multiscale transport & Finite-volume kinetic schemes & CPU, GPU; related distributed tools & Mesoscopic modeling plus SciML/differentiable workflows \citep{Xiao2021Kinetic} \\
\addlinespace
\resizebox{2.2cm}{\height}{\pkg{TrixiParticles.jl}} & Particle-based multiphysics & SPH and particle methods & CPU/GPU-oriented ecosystem & Particle multiphysics within the broader Trixi family \citep{Neher2025TrixiParticles} \\
\addlinespace
\pkg{LCS.jl} & Turbulent particle-laden flow & Eulerian fluid + Lagrangian particles & CPU/GPU, multi-GPU & Single-source multiphase HPC and GPU-native particle communication \citep{LCS2026} \\
\bottomrule
\end{tabularx}
\end{table}

\subsection{Compressible and high-order solvers}

Trixi.jl provides a high-order framework for hyperbolic PDEs in Julia. Its equations, numerical fluxes, meshes, semidiscretizations, time-integration algorithms, and diagnostics are organized through generic interfaces, with support for entropy-stable formulations, shock capturing, and adaptive meshes \citep{Ranocha2022Trixi}. The architectural payoff is visible when a major algorithm changes: TrixiLW.jl introduces Lax--Wendroff flux reconstruction while reusing curved-mesh and analysis infrastructure through an ordinary package dependency \citep{Babbar2024TrixiLW}. H3PC and MARUT test the same foundation under hypersonic thermochemistry, nonconforming adaptivity, and accelerator-oriented reacting-flow requirements \citep{Peyvan2025H3PC,Mondal2026MARUT}.

This family is strong evidence that specialization and reuse can coexist in a performance-oriented framework. High-order element-local arithmetic is, however, comparatively well aligned with compiler specialization and local kernels; arbitrary polyhedral flow adds further challenges in geometry, data access, and solver structure. The published scaling results are assessed in \cref{sec:performance}. Here the narrower conclusion is architectural: Julia can support extensions that alter equations or discretizations without forcing each research code to reproduce the surrounding solver infrastructure.

\subsection{Incompressible flow and differentiable turbulence}

WaterLily.jl and IncompressibleNavierStokes.jl make accelerator portability and differentiation part of the solver architecture on regular Cartesian grids. WaterLily combines an immersed-boundary formulation for moving bodies with compact, multidimensional kernels that can target CPUs or GPUs and participate in differentiated computations \citep{Weymouth2025WaterLily}. IncompressibleNavierStokes.jl uses energy-conserving staggered discretizations, matrix-free kernels, and handwritten discrete adjoints to support turbulence simulation and solver-in-the-loop closure training \citep{Agdestein2026INS}. Their different strategies show how a differentiable interface can combine automatic, custom, and handwritten derivative implementations.

The regularity that makes these designs effective also limits the inference that can be drawn from them. Fixed topology and predictable stencil access favor generic arrays, accelerator kernels, and reverse-mode differentiation, whereas polyhedral connectivity, changing meshes, and strongly coupled physics introduce additional data and solver dependencies. These packages therefore mark the region in which Julia's single-source model is presently most convincing; their quantitative GPU and distributed-memory evidence is considered separately in \cref{sec:performance}.

\subsection{Unstructured finite-volume and finite-element solvers}

Unstructured engineering CFD places different demands on performance portability because mesh connectivity, boundary treatment, and pressure--velocity coupling introduce indirect access and problem-dependent sparse structure. XCALibre.jl addresses this setting with cell-centered finite-volume formulations on structured and unstructured meshes and an XPU-oriented implementation \citep{Medina2025XCALibre}. Its face-based operations must gather states through connectivity arrays, accumulate contributions in cells, and incorporate boundary and partition information. On accelerators, these operations produce irregular reads and potential write conflicts, so CPU and GPU implementations may require different traversal or accumulation strategies even when they share the same flux and discretization interfaces.

Pressure-based formulations also depend strongly on linear algebra. Their systems contain null spaces, pressure--velocity block structure, and preconditioning requirements that a generic sparse interface cannot infer from the matrix alone. Assembled matrices provide access to established algebraic solvers, whereas matrix-free operators reduce storage but must supply efficient operator actions and suitable approximations for preconditioning. The Gridap family illustrates another route to organizing this structure: Gridap.jl represents weak forms and finite-element spaces, GridapSolvers.jl provides distributed solver layers, and SegregatedVMSSolver.jl constructs an LES application from these components \citep{Badia2020Gridap,Manyer2024GridapSolvers,Brunelli2025SegregatedVMS}. Here, reuse occurs through variational forms, distributed data, and solver interfaces rather than through the flux-level abstractions used by Trixi.

Together, these projects show that Julia's compositional model extends to unstructured operators and application-level flow algorithms, but the quantitative evidence is still thinner than for structured-grid solvers. Controlled studies rarely separate residual evaluation, matrix construction, linear solution, preconditioning, and communication across multiple backends. Reporting these costs, together with the amount of backend-specific code, would make it possible to distinguish reusable software structure from performance demonstrated only for one mesh or device. The remaining industrial workflow gaps are discussed in \cref{sec:limitations}.

\subsection{Geophysical, ocean, and atmospheric models}

Geophysical projects test Julia under long integrations, large data volumes, and complex parameterizations. Oceananigans.jl implements hydrostatic and nonhydrostatic flow in a GPU-oriented environment that spans idealized studies and distributed ocean configurations \citep{Ramadhan2020Oceananigans,Wagner2025Oceananigans}. SpeedyWeather.jl emphasizes an interactive and extensible atmospheric model based on spectral methods \citep{Klower2024SpeedyWeather}. Together they show that high-level access can serve both specialized accelerator execution and rapid modification of a complete dynamical core.

CliMA extends the architectural test from a fluid solver to a climate-model system. Its Julia components combine resolved atmospheric dynamics with physical parameterizations, calibration, and data-assimilation workflows \citep{Sridhar2022ClimateMachine,Yatunin2026ClimaAtmos}. This form of composition is scientifically different from conventional engineering multiphysics but structurally similar: components with different numerical behavior must share state, execution resources, and verification practices. The resulting quality criteria include conservation over long time scales, restart and I/O reliability, and reproducible experiment configuration in addition to kernel speed.

\subsection{Kinetic, particle, and multiphase methods}

Kinetic and particle methods enlarge both the state space and the communication problem. Kinetic.jl places both continuum and Boltzmann-type descriptions in a common finite-volume and scientific-machine-learning setting, allowing transport models and uncertain closures to participate in shared numerical workflows \citep{Xiao2021Kinetic}. TrixiParticles.jl explores reuse across a package family even when the representation changes from element-based PDEs to particle methods \citep{Neher2025TrixiParticles}. These cases test composability across modeling formalisms as well as within a single discretization.

LCS.jl provides the complementary systems case: Eulerian fluid fields, Lagrangian particles, device-resident sorting, and inter-device migration must operate together \citep{LCS2026}. Its performance evidence is summarized in \cref{sec:performance}; architecturally, it shows why source portability alone is insufficient for multiphase HPC. Particle containers, continuum fields, and distributed decompositions have different ownership and migration semantics, so effective composition still requires explicit control of memory placement, synchronization, and communication.

\subsection{Cross-cutting infrastructure and ecosystem composition}

Julia CFD depends heavily on infrastructure developed outside fluid mechanics. MPI.jl and the GPU ecosystem provide execution mechanisms; DifferentialEquations.jl and NonlinearSolve.jl supply reusable solver interfaces; and differentiation packages offer complementary derivative strategies \citep{Byrne2021MPI,Besard2019GPU,Rackauckas2017DifferentialEquations,Pal2024NonlinearSolve,Revels2016ForwardDiff,Moses2020Enzyme,Dalle2025DifferentiationInterface}. Their value comes from a shared language-level interface through which arrays, model functions, and numerical types can cross package boundaries without first being reduced to a foreign-function ABI.

Composition nevertheless creates its own compatibility surface. A valid CPU operation may allocate on the host when used inside a GPU workflow, while a numerically correct primitive may lack a derivative rule or distributed implementation. Ecosystem maturity therefore depends on whether components can be substituted while preserving correctness, performance, and reproducibility. Current evidence consists of several partial demonstrations of that property; a unified CFD platform has yet to emerge.

\section{Beyond the forward solver}
\label{sec:beyond_forward}

The architectural argument for Julia becomes most consequential when CFD is embedded inside a larger computational task. A forward solver can tolerate substantial separation between components because data may cross an interface only at initialization and output. Optimization, uncertainty quantification (UQ), and machine learning change this pattern. They repeatedly call the solver, alter parameters or models, require derivatives, and often execute learned components inside the time-stepping loop. This section therefore focuses on workflows in which simulation functions as an inner algorithm within a larger computation.

\subsection{Sensitivity analysis, inverse problems, optimization, and uncertainty}

Consider a semi-discrete CFD model
\begin{equation}
  \frac{d\mathbf{q}}{dt}=\mathbf{R}(\mathbf{q},\boldsymbol{\theta}),
  \qquad
  J = J\!\left(\mathbf{q}(t;\boldsymbol{\theta}),\boldsymbol{\theta}\right),
\end{equation}
where $\boldsymbol{\theta}$ denotes design variables, uncertain parameters, or controls. The forward calculation maps these quantities to a trajectory or steady state, whereas optimization and inference also need derivatives of objectives or observations. Forward sensitivities are attractive for few parameters; adjoint methods are generally preferable when a small number of objectives depends on a large parameter space. This distinction is algorithmic and remains independent of the language used to implement it.

Automatic differentiation changes how derivative operations are produced, not the mathematics of sensitivity analysis. Local Jacobian products may be generated with ForwardDiff, Zygote, Enzyme, or another backend, while the outer method remains a discrete adjoint, an implicit derivative of a converged solve, or a checkpointed reverse pass through time integration \citep{Revels2016ForwardDiff,Moses2020Enzyme,Dalle2025DifferentiationInterface}. Such hybrid strategies are important because CFD resists undifferentiated black-box treatment: long trajectories create storage--recomputation tradeoffs, iterative solves require a choice between algorithmic and implicit derivatives, and shocks or changing mesh topology introduce nonsmoothness. Communication and accelerator primitives must also participate in the derivative program, while chaotic trajectories may make instantaneous long-time sensitivities scientifically inappropriate even when they are computable.

The Julia examples accordingly use different derivative levels. IncompressibleNavierStokes.jl supplies handwritten adjoints behind a differentiable solver interface, WaterLily.jl admits broad AD use in a compact immersed-boundary code, and Arrhenius.jl differentiates combustion-kinetics workflows \citep{Agdestein2026INS,Weymouth2025WaterLily,Ji2021Arrhenius}. Their common contribution is the treatment of sensitivity as a normal solver capability, with specialized derivatives where necessary and common interfaces to optimization or inference.

UQ introduces a related but distinct set of compositions. Non-intrusive sampling and ensemble Kalman methods need many solver instances whose parameter values and failure modes vary; Bayesian inference repeatedly couples simulation outputs to likelihood evaluation; intrusive polynomial or interval methods may ask numerical operators to accept scalar types richer than ordinary floating point. Julia's generic numerical and ensemble infrastructure can reduce the need for a special UQ-specific version of the flow code. This is a plausible architectural advantage, particularly when the same parameter container and observation operator can be shared by simulation, differentiation, and inference.

CFD-specific evidence for that advantage remains limited. The surveyed literature contains stronger demonstrations of Julia for forward HPC and local or trajectory-level differentiation than for end-to-end uncertainty propagation in three-dimensional flow. Robustness under parameter variation, sample reduction, multifidelity control variates, load balancing for unequal trajectory costs, and treatment of model-form discrepancy remain application problems. Type genericity helps only if the discretization and its dependencies genuinely support the new scalar or ensemble representation; otherwise the apparent generality ends at a sparse library, device kernel, or external model.

Design, calibration, control, and UQ also change the appropriate unit of performance. Three-dimensional inverse workflows depend on robust solves across parameter variation, accurate gradients, checkpointing, reduced models, and efficient ensemble scheduling. Compilation may be negligible for one long simulation but significant when thousands of short trajectories start independently, while a fast gradient is of little value if it is inaccurate enough to mislead the outer algorithm. The relevant measures are therefore time to a converged design, posterior, or calibrated model and the reliability of that result, not seconds per isolated CFD step. This gap between available infrastructure and application-level evidence is one reason complete inverse workflows should become a priority for future benchmarks.

\subsection{Physics--machine-learning hybrids and solver-in-the-loop learning}

Machine learning can enter CFD as an offline surrogate, an embedded closure, or a component trained through the time-integrated solver. Only the latter two strongly test composability because the learned function must share state, devices, and numerical constraints with the flow code. The universal-differential-equation viewpoint captures this arrangement: trusted transport and geometric structure remain mechanistic while selected uncertain terms become trainable \citep{Rackauckas2020UDE}. This division keeps known physical and numerical structure explicit while concentrating learning on uncertain components.

Julia can represent such a learned term as an ordinary callable component operating on the same arrays as the surrounding discretization. Kinetic.jl combines kinetic modeling with scientific-machine-learning workflows, while IncompressibleNavierStokes.jl trains neural closures a posteriori inside LES trajectories \citep{Xiao2021Kinetic,Agdestein2026INS}. Avoiding a mandatory conversion to a separate tensor runtime matters most when a relatively small learned model is invoked at every grid point or time step. Solver-in-the-loop training then tests the whole vertical stack: physics, kernels, differentiation, optimization, and accelerator execution must remain compatible.

Deep integration also exposes constraints that offline fitting can conceal. A learned closure may destabilize the PDE or violate conservation, positivity, symmetry, or entropy properties even when it matches training data. Generic model interfaces make alternative placements and parameterizations easier to express, but they do not supply the required numerical analysis. The surrounding workflow may still use external ML, meshing, chemistry, and vendor libraries while retaining a programmable solver architecture. What matters is whether those components can participate without reducing the solver to a black-box executable.

\section{Performance and scalability: from language claims to CFD evidence}
\label{sec:performance}

Language shootouts are a poor basis for assessing CFD software. A result that one kernel is ``as fast as C'' says little about mesh traversal, sparse algebra, pressure solves, communication, compilation, I/O, or the cost of embedding the solver in an inverse problem. A more informative question is whether published Julia applications have reached scales and architectures that are relevant to contemporary CFD. \Cref{tab:perf_evidence} organizes such evidence by the capability demonstrated in each study.

\begin{table}[t]
\centering
\caption{Representative published scale and performance evidence for Julia fluid and HPC software. Each row records a demonstrated capability together with the hardware and computational setting reported in the source study.}
\label{tab:perf_evidence}
\scriptsize
\setlength{\tabcolsep}{4pt}
\renewcommand{\arraystretch}{1.13}
\begin{tabularx}{\textwidth}{>{\raggedright\arraybackslash}p{2.6cm} >{\raggedright\arraybackslash}p{4.0cm} >{\raggedright\arraybackslash}p{3.0cm} X}
\toprule
\textbf{System} & \textbf{Reported scale / result} & \textbf{Architecture} & \textbf{What the result establishes} \\
\midrule
\pkg{Trixi.jl} & Simulations on up to 61,440 CPU cores; comparison with Fortran FLUXO & Distributed CPU / MPI & High-order Julia CFD can operate at conventional leadership-class CPU concurrency; startup/JIT costs remain visible \citep{Candelaresi2026MassivelyParallel}. \\
\addlinespace
\pkg{WaterLily.jl} & Inter-node weak-scaling efficiency $>96\%$ to one billion cells in reported tests & CPU/GPU-capable MPI solver & Backend-agnostic structured CFD can extend from single-device use to large distributed grids \citep{Font2026WaterLilyScaling}. \\
\addlinespace
\pkg{Oceananigans.jl} & 488 m global-ocean simulation on 768 NVIDIA A100 GPUs & Distributed GPU & Julia-native geophysical CFD can sustain global high-resolution workloads on hundreds of accelerators \citep{Silvestri2023OceananigansScaling}. \\
\addlinespace
\resizebox{2.45cm}{\height}{\pkg{IncompressibleNavierStokes.jl}} & Double-precision DNS to $840^3$ on one GPU & Single GPU & Differentiability and memory-aware design can coexist with substantial turbulence resolutions \citep{Agdestein2026INS}. \\
\addlinespace
\pkg{LCS.jl} & $>85\%$ strong efficiency to 256 GPUs; $>90\%$ weak efficiency to 216 GPUs; reported max 18$\times$ GPU/CPU speedup & Multi-GPU particle-laden DNS & Extends single-source portability to GPU-native particle communication \citep{LCS2026}. \\
\addlinespace
Frontier workflow study & Weak scaling to 4096 MPI processes/GPUs on 512 nodes; near-zero binding overhead but about 50\% kernel gap to native HIP in the tested stencil & AMD MI250X / exascale system & Julia's end-to-end HPC model is feasible, while performance portability still requires backend-specific scrutiny \citep{Godoy2023Frontier}. \\
\bottomrule
\end{tabularx}
\end{table}

\subsection{Scope of the published performance evidence}

The studies in \cref{tab:perf_evidence} document Julia applications at substantial distributed CPU and accelerator scales. They test whether Julia's abstractions survive communication, memory pressure, compilation, solver complexity, and workflow composition. Because the evidence spans several discretizations and communication patterns, it is stronger than a collection of microbenchmarks, although coverage remains concentrated in research applications.

The rows cannot be normalized into a speed ranking: their equations, algorithms, hardware, precision, and definitions of efficiency differ. Comparisons with lower-level implementations are likewise conditional on data layout, external libraries, compiler quality, and developer effort. The table therefore establishes demonstrated capability, not a language-wide performance ratio. It also makes negative evidence relevant: a binding layer can be inexpensive while a portable kernel still trails a native backend, so each layer must be measured on its own terms.

\subsection{Performance portability across single devices and distributed systems}

Specialization removes interpretation from hot loops, but it does not remove hardware constraints. Regular-grid solvers can often move substantial numerical source across CPUs and GPUs because their access patterns and temporary storage are predictable \citep{Ramadhan2020Oceananigans,Weymouth2025WaterLily,Agdestein2026INS}. Portable reductions, scans, and other primitives extend this model across accelerator vendors \citep{Nicusan2025AcceleratedKernels,Pilliat2026KernelForge}. Their performance still depends on memory layout, vectorization or coalescing, register use, launch granularity, and the quality of the lowest-level primitive selected by the backend.

Irregular algorithms require a looser meaning of ``single source.'' Unstructured finite-volume traversal, adaptive topology, and particle migration may need different layouts or kernel decompositions on CPUs and GPUs even when their mathematical interface is shared \citep{Medina2025XCALibre,LCS2026}. Dispatch can preserve common solver logic while selecting specialized implementations. Here, portability is measured by retained performance and the amount of backend-specific code required, not by textual identity of the kernels.

Distributed execution adds decomposition, load balance, communication, and I/O to this accounting. MPI.jl can use system MPI with low binding overhead, but application scaling still depends on the communication structure chosen by the solver \citep{Byrne2021MPI}. The Frontier study makes the separation explicit: Julia bindings and parallel I/O scaled well while the tested generated stencil remained slower than native HIP \citep{Godoy2023Frontier}. Julia's strongest portability claim is consequently two-dimensional: numerical source should remain shared where appropriate, and modest specialization behind stable interfaces should retain competitive performance across architectures.

\subsection{Compilation, startup, and end-to-end measurement}

Just-in-time specialization makes workload definition part of performance measurement. Package loading and the first invocation incur one-time costs that later calls avoid; they may be negligible in a long simulation but dominant in a short ensemble member. At scale, simultaneous loading and compilation can also stress shared filesystems, as observed in the Trixi study \citep{Candelaresi2026MassivelyParallel}. Precompiled caches or application images can reduce this cost and should be included in the documented deployment procedure and performance accounting.

Benchmarks should therefore separate time to load, one-time specialization, steady-state throughput, and complete workflow time, while also reporting memory and cache requirements. The appropriate balance changes between a persistent production run and an optimization or uncertainty-quantification campaign that starts many short trajectories. Ahead-of-time C++ or Fortran applications incur analogous build work before job launch; a fair comparison should report these costs at whichever stage they occur.

The architectural claim also concerns engineering effort. A portable implementation that retains most architecture-specific performance with little duplicate source may be preferable to a faster code that requires independent kernels for every backend, whereas an expanding collection of exceptions would erase that advantage. Runtime evidence should therefore be paired, where possible, with the amount of backend- and derivative-specific code and with end-to-end measures such as time to a verified gradient or converged inverse result.

\section{Julia relative to adjacent CFD software paradigms}
\label{sec:comparison}

A language-centered survey can easily confuse ecosystem trends with language-specific advantages. Modern CFD software is converging toward heterogeneous execution, higher-level abstractions, code generation, differentiable programming, and stronger separation between models and numerical infrastructure. Julia represents one implementation of this broader trend. The most useful comparison is therefore with contemporary software architectures that solve similar integration problems in different ways.

\subsection{Mature C++ performance-portability and solver ecosystems}

C++ frameworks provide an established baseline for generic abstractions and heterogeneous execution. AMReX combines block-structured AMR, parallel data structures, and large-scale CPU/GPU execution in a mature ecosystem \citep{Zhang2019AMReX}; related frameworks use templates, execution policies, and backend-specific kernels while retaining established MPI, solver, and mesh infrastructure. Predictable deployment artifacts and organizational continuity are decisive advantages for long-lived production codes. These modern architectures provide the relevant comparison for Julia.

AMReX also illustrates the value of concentrating effort in a stable infrastructure layer. Applications inherit mesh hierarchy management, load balancing, communication, particle containers, I/O, and device execution from a framework tested by several large code bases. This shared substrate limits freedom at some interfaces but amortizes difficult systems engineering over a much larger community. Julia's package modularity distributes responsibility more finely; that can make numerical components easier to replace, but it does not yet provide the same depth of common application infrastructure or institutional validation.

The difference is primarily where composition occurs. C++ commonly fixes more structure through templates and a distinct build phase, whereas Julia keeps types and generic functions directly accessible during interactive development and specializes their concrete combinations at execution. This can shorten experiments with new equations, scalar types, or differentiated components, at the cost of runtime compilation and a younger deployment toolchain. Both approaches can reach the hardware; Julia is attractive when changing and recombining the software is itself a central part of the workload.

The comparison should include organizational cost as well as language mechanics. A C++ team may accept a longer edit--compile cycle because it has mature debuggers, performance tools, release procedures, and personnel familiar with the framework. A Julia project may gain faster method development but take responsibility for deployment recipes, package compatibility, and accelerator edge cases that the established framework has already absorbed. The balance therefore changes with project age, team expertise, and how often the numerical architecture itself must change.

\subsection{Domain-specific languages, code generation, and differentiable array systems}

Domain-specific languages gain optimization power by restricting what the user can express. Firedrake, for example, transforms high-level finite-element weak forms into lower-level kernels while exploiting variational structure unavailable to a general-purpose compiler \citep{Rathgeber2016Firedrake}. This is effective inside the domain, but unusual data structures, particle algorithms, or external solver loops may cross into another programming layer. Julia uses a more permeable boundary: packages can introduce symbolic forms or generated kernels where global transformation helps while the surrounding workflow remains an ordinary program.

Firedrake's restriction also enables system-wide numerical services: a weak form can participate in automated assembly, differentiation, boundary treatment, and solver configuration because its mathematical meaning is available to the transformation system. Julia's ordinary functions are easier to mix with arbitrary code, but their semantics are not automatically visible to the compiler or a solver package. A Julia DSL can recover this information locally, yet doing so introduces the same design question faced by other DSLs: which operations remain analyzable, and what happens when a user escapes into general code? Host-language flexibility and domain knowledge are complementary rather than interchangeable advantages.

JAX-based CFD provides a closer comparison for differentiable accelerator computing. JAX-CFD combines finite-volume and pseudospectral solvers with automatic differentiation and learned turbulence models in a functional array framework \citep{Kochkov2021JAXCFD}. JAX-Fluids 1.0 and 2.0 extend this pattern to differentiable compressible multiphase flow and large accelerator runs; the latter's reported scaling to hundreds of GPUs and more than a thousand TPU cores shows that transformation-based array systems are not confined to small ML-adjacent demonstrations \citep{Bezgin2025JAXFluids2}. Julia instead attempts to transform a wider language surface that includes mutation, custom scalar and container types, sparse data, and conventional scientific libraries. That flexibility accommodates more existing programming patterns but produces a heterogeneous derivative stack in which generated, custom, and implicit rules may coexist.

The functional array model imposes constraints, but those constraints make compilation boundaries and program transformations comparatively coherent. Vectorization, differentiation, and device placement operate on a representation built for whole-program analysis, and JAX benefits from a large adjacent machine-learning ecosystem. Julia is less uniform: it can incorporate irregular data structures and existing libraries more naturally, while AD or accelerator compatibility may fail at a particular mutation, external call, or unsupported primitive. The practical choice is thus not ``general language versus restricted language'' in the abstract, but whether the dominant CFD workload fits the representation well enough that its transformation guarantees outweigh the cost of adapting algorithms to it.

The tradeoff is thus between the optimization leverage of constrained representations and the reach of a general composition language. DSLs and JAX can offer exceptionally coherent transformations for algorithms that fit their models; C++ provides mature systems infrastructure; Julia aims to let specialized representations coexist without defining the whole application boundary. The relevant choice depends on which parts of the workflow must remain open to change.

\subsection{Julia's architectural distinction}

Julia's architectural distinction lies in its integration point: numerical models, device arrays, derivative rules, and outer algorithms can meet through the host language's types and generic functions. When their interfaces align, a change can propagate from high-level physics into specialized execution without first creating a separate application boundary. GPU, MPI, AD, and machine-learning capabilities are also available in other ecosystems, and performance parity with hand-tuned implementations remains workload dependent \citep{Godoy2023Frontier}.

That advantage is presently stronger vertically within designed package families than horizontally between unrelated solvers. A shared language does not reconcile mesh representations, boundary semantics, device assumptions, or derivative conventions by itself. The central empirical question is therefore whether Julia reduces the \emph{total engineering cost} of evolving a code---including backend-specific source, verification effort, and performance lost to genericity---relative to C++ frameworks, DSLs, or transformation-based array systems. This comparison concerns software architecture rather than syntax.

\section{Software engineering, verification, and reproducibility}
\label{sec:software_engineering}

The sustainability of CFD software depends on properties that are difficult to infer from performance plots: reproducible environments, verification tests, API stability, documentation, and the ability to move from an exploratory notebook to a production job without silently changing the numerical method. Julia's package model and interactive development style provide useful tools for this process, but they also create a large combinatorial space of package and backend configurations that must be managed deliberately.

\subsection{Reproducible environments and layered numerical testing}

Julia environments can record direct and transitive package versions and pin binary artifacts, so a paper can archive its solver, dependencies, inputs, and analysis scripts together. This captures more than a solver tag, but it does not by itself ensure numerical reproducibility. Hardware, precision, decomposition, tolerances, initialization, stochastic settings, and reduction order may change a result even when the manifest is identical. Reproducible CFD therefore requires the software environment and the numerical experiment description.

It is useful to distinguish three targets. \emph{Repeatability} asks whether the same environment on the same platform reproduces a run; \emph{reproducibility} asks whether an independent environment obtains scientifically consistent results; and \emph{replicability} asks whether an independent implementation supports the same conclusion. A Julia manifest addresses the first target most directly. It cannot preserve an unavailable system MPI, guarantee bitwise agreement after a GPU compiler changes, or show that a conclusion is insensitive to a particular implementation. Archiving containers or application images, raw configuration files, and representative outputs can extend the record, but numerical tolerances and expected invariants must still be stated.

Parallel fluid simulation makes bitwise identity an especially fragile standard. Reduction order changes with domain decomposition, thread scheduling, and accelerator kernels; chaotic trajectories can separate rapidly while retaining consistent statistical behavior; and stochastic forcing or particle models require explicit control of random-number streams across processes and devices. A reproducibility package should therefore define what agreement means for the application. Conservation errors, spectra, integrated forces, ensemble moments, convergence rates, or distributions may be more meaningful targets than equality of every state value.

Generic components make layered testing natural. Fluxes and boundary operators can be checked independently, semidiscretizations against manufactured solutions, and accelerator kernels against CPU references; alternative precision or number types can reveal hidden assumptions. The governing criteria remain numerical rather than language specific: conservation, convergence, preservation properties, stability, and gradient fidelity matter more than source-level test coverage. These checks become more valuable as equations, meshes, integrators, devices, and derivative systems are recombined.

Verification should follow the dependency graph of the numerical method. Local properties such as consistency, symmetry, positivity, or discrete entropy behavior can be checked before a complete flow case obscures the source of an error. Manufactured solutions and grid-convergence studies then test the assembled spatial and temporal method, while canonical flows provide validation against accepted physical evidence. Differentiable solvers add another layer: Jacobian actions and adjoints should be compared with finite differences or independent derivations, and the resulting optimization trajectory should be checked for sensitivity to derivative tolerances. Passing an ordinary forward regression test does not establish gradient correctness.

The same combination space creates a maintenance burden. Continuous integration cannot exercise every runtime, accelerator, MPI implementation, and model, so projects need a declared support matrix. Small semantic components can be tested broadly, representative solver configurations continuously, and expensive scale or regression cases periodically on supported hardware. Composability replaces some duplicated implementation work with compatibility-contract work; it does not eliminate it.

A useful support matrix should distinguish code paths that are merely type generic from combinations receiving routine validation. For example, a solver may compile for several scalar types but test production cases only in two precisions, or expose several GPU backends while regularly benchmarking one. Stating this boundary makes extensibility credible because users can separate an intended extension point from a promised production configuration. Periodic performance tests are also necessary: numerical correctness can remain unchanged while a new compiler or dependency introduces allocations, excess communication, or a large startup regression.

\subsection{From interactive research code to production deployment and long-term maintenance}

Julia's interactive environment shortens the path from a new numerical idea to a packaged component because the same functions can be inspected, modified, and subsequently deployed without translation. Production use nevertheless requires a more fixed environment: curated manifests, precompiled images, artifact caches, explicit runtime versions, and reliable restart and parallel I/O. Large-scale studies show that this transition is feasible and that successful production applications increasingly use deliberately assembled software images.

Supercomputing centers impose additional constraints that are largely invisible in workstation development. Compute nodes may lack network access, system MPI and accelerator stacks must match site policy, shared filesystems penalize many small package files, and jobs may need to restart after queue limits or hardware failures. Production workflows therefore require offline-deployable artifacts, predictable startup, compact provenance records, scalable checkpointing, and logs that identify the precise software image. These requirements should be designed and tested early in a Julia application's development.

The software image can combine Julia with established C or Fortran infrastructure and specialized C++ libraries; H3PC's use of Mutation++ is one example \citep{Peyvan2025H3PC}. This is often the practical route to mature meshing, partitioning, I/O, or thermochemistry. Long-term sustainability then depends on governance---maintainers, releases, documentation, regression tests, and stable interfaces---more than on syntax. Shared infrastructure distributes some work across communities, but the concentration of important packages in small teams remains a material risk.

Foreign libraries nevertheless reintroduce boundaries that must be managed explicitly. Array ownership, device residency, threading and MPI initialization, error handling, and binary compatibility can all affect correctness or performance even when the call overhead is negligible. A robust Julia wrapper should expose the semantic capability of the library rather than mirror every low-level entry point, preserve zero-copy paths where possible, and include integration tests against the versions actually deployed. Interoperability is most valuable when it localizes a boundary instead of allowing external assumptions to spread through the solver.

Governance ultimately determines whether these practices survive the original research project. Stable release archives and citation metadata make results traceable; deprecation policies give downstream packages time to adapt; contributor documentation reduces dependence on one expert; and benchmark ownership prevents silent decay of performance claims. Julia's modular ecosystem can pool such work in shared MPI, GPU, AD, and solver packages, but it can also propagate an upstream API change across many applications. Production readiness therefore depends on both local project discipline and the health of the infrastructure on which the project depends.

\section{Current limitations and gaps}
\label{sec:limitations}

The preceding examples establish Julia's capability for serious fluid simulation while also showing a substantial gap from the breadth of established CFD environments. Those environments have accumulated decades of physical models, meshing workflows, validation databases, user support, and operational knowledge. Julia is strongest today in research-facing numerical software and selected large-scale applications. Its limitations can be grouped into three broad categories: ecosystem breadth and industrial workflow integration; deployment and heterogeneous-performance maturity; and algorithmic limits of differentiable/composable execution.

\subsection{Ecosystem breadth, meshing, and industrial workflow integration}

Julia's CFD community is small relative to established open-source and commercial environments, and the difference is qualitative as well as numerical. Mature ecosystems accumulate difficult cases, robustness heuristics, validation databases, support knowledge, and complete preprocessing and postprocessing practices. Julia has capable solver-side components and can consume established mesh formats, but it lacks a comparably integrated path from complex geometry through production case management. Likewise, its coverage of industrial turbulence, reacting, rotating, and multiphase workflows remains fragmented despite rapid progress in research frameworks such as H3PC and MARUT \citep{Medina2025XCALibre,Peyvan2025H3PC,Mondal2026MARUT}. Reliable interoperability is therefore a more credible near-term strategy than reproducing every mature tool natively.

Fragmentation also appears within the Julia ecosystem: projects define different mesh, field, boundary, communication, and output semantics. Forcing a universal representation would erase legitimate differences among discretizations, but leaving every boundary private prevents infrastructure reuse. The useful target is semantic interoperability for selected operations---geometry queries, ownership, operator actions, field metadata, and standard output---with specialized internal layouts preserved. This is the principal gap between the vertical composability already demonstrated inside projects and horizontal composition across them.

\subsection{Compilation, deployment, and uneven performance portability}

Compilation latency is most visible in short jobs and large launches that amplify file-system and code-loading costs \citep{Candelaresi2026MassivelyParallel}. Curated images and caches mitigate it but add build and deployment discipline. Performance portability is also uneven: regular stencils and element-local kernels map more readily across devices than sparse gathers, adaptive meshes, searches, or particle migration. The observed gap between a generated Julia stencil and native HIP illustrates why a program running on several vendors is not enough evidence that it performs well on them \citep{Godoy2023Frontier}.

High-level portability also depends on low-level ecosystem coverage. Sparse solvers, multigrid, FFTs, partitioners, and collectives are not equally mature on every accelerator, and an abstract interface cannot supply a missing implementation. Portable primitives help close this gap \citep{Nicusan2025AcceleratedKernels,Pilliat2026KernelForge}, while profiling and kernel inspection are needed to reveal when generic source produces allocations or poor machine code. Accessibility without performance transparency would merely move complexity out of sight.

\subsection{Differentiability, multiphysics composition, and algorithmic limits}

Rich AD tooling does not make realistic CFD uniformly differentiable. A local flux accepting dual numbers, a reverse-mode time step, an implicit steady adjoint, and a statistically meaningful turbulent sensitivity are distinct capability levels. Memory, iterative algorithms, discontinuities, topology changes, external libraries, and communication complicate the software, while chaos and bifurcation can invalidate a mechanically produced derivative as the desired scientific quantity. Packages should state the derivative operation and objective they support rather than use ``AD compatible'' as a binary label.

Multiphysics composition has the same distinction between interface and algorithm. A common language simplifies calls and data exchange, but conservative transfer, synchronized integration, nonlinear convergence, ownership, and load balance determine whether the coupled method is valid and scalable. Julia lacks widely adopted coupling interfaces that have been validated on distributed heterogeneous systems. Moreover, every added mesh, backend, derivative method, or closure expands the untestable Cartesian product of configurations; production projects must distinguish generic extensibility from combinations they actually verify and support.

\section{Emerging directions}
\label{sec:future}

The most promising future directions are areas in which Julia's combination of generic numerics, heterogeneous execution, differentiation, and scientific-machine-learning infrastructure could change how CFD algorithms are assembled. Three directions appear particularly consequential: sensitivity-aware and AI-native solver design; interoperable multiscale/multiphysics components; and compiler-assisted numerical software evaluated with end-to-end benchmarks.

\subsection{Sensitivity-aware and AI-native CFD}

A sensitivity-aware solver would design derivative actions alongside the forward operators rather than attach an adjoint after the data structures and algorithms are fixed. Linear and nonlinear components would expose Jacobian and transpose actions, time integrators would define checkpointing semantics, and boundary or closure models would state their derivative contracts. Julia can prototype this architecture with local AD, implicit rules, and custom adjoints coexisting behind common functions. The standardization target should be the promised mathematical action, not one derivative technology.

The same interfaces allow learned components to participate inside the solver rather than only consume files produced by it. Such ``AI-native'' use is valuable when closures, controllers, or reduced models must be executed and differentiated within a trajectory. Its scientific requirement is structure: conservation, positivity, invariance, realizability, and stability constraints should be explicit in component APIs rather than left to an arbitrary black-box function. Composability is useful because it permits these constraints to remain part of the numerical skeleton.

High-fidelity and learned models are consequently likely to form adaptive workflows rather than compete as substitutes. A solver may generate targeted data, correct a reduced model, estimate its error, or resume high-fidelity resolution when the surrogate leaves its training regime. These applications amplify the need for verified derivatives and uncertainty estimates and make end-to-end behavior more important than the throughput of either component in isolation.

\subsection{Composable multiphysics, multiscale models, and common interfaces}

Multiscale flow couples representations with different state dimensions, time scales, and ownership rules. Kinetic--continuum, particle--fluid, and fluid--structure systems therefore need conservative projection and exchange, synchronized integration, nonlinear coupling, and distributed load balance. Julia can express model selection and callbacks conveniently, but these algorithmic contracts determine whether the composition is scientifically and computationally credible.

Progress requires semantic interfaces rather than a universal data layout. Packages could agree on operations for geometry, owned and ghost entities, field transfer, operator actions, and output while retaining specialized storage for structured grids, adaptive trees, unstructured cells, or particles. Mature external libraries for partitioning, solvers, I/O, chemistry, and meshing should remain part of this architecture; high-level workflow ownership does not require reimplementation or language purity.

Common scientific interfaces would also create shared verification and benchmarking infrastructure. Manufactured solutions, error measures, transfer tests, and partitioning experiments could then be applied across discretization families without forcing those families into a single solver. This narrower layer of interoperability is more attainable, and potentially more valuable, than a unified Julia representation for all CFD data.

\subsection{Compiler-assisted numerical software and end-to-end benchmarking}

Compiler specialization could use numerical structure more directly: polynomial order, equation type, precision, and device can guide fused kernels, local thermochemical solvers, stencil construction, or matrix-free derivative actions. Portable Julia primitives already show that generic interfaces can approach tuned implementations when their abstractions reflect hardware behavior \citep{Carrica2025TRMM,Pilliat2026KernelForge,Nicusan2025AcceleratedKernels}. CFD packages can extend this idea by exposing domain information to transformations without making the entire application a separate DSL.

The design must remain inspectable and maintainable. Generated implementations should sit behind stable extension points, retain transparent fallbacks, and be checked against the generic numerical algorithm. Otherwise compiler assistance merely replaces duplicated source with fragile hidden machinery. Progressive specialization---starting from a correct reference operation and adding verified optimized paths---is a more sustainable model than requiring solver authors to understand compiler internals.

Evaluation should follow the complete claim. In addition to throughput and scaling, studies should report startup, memory and communication, cross-vendor behavior, gradient accuracy and cost, full inverse-workflow time, reproducibility, and the amount of specialized source maintained. Productivity cannot be reduced reliably to lines of code, but case studies can record the effort to add a model or backend and the fraction of tests and operators reused. These measures test the consequential question: whether an architecture remains expressive, portable, differentiable, and maintainable as its scientific role changes.

\section{Discussion}
\label{sec:discussion}

The evidence supports a qualified maturity claim. Julia fluid software now spans a range of numerical and application settings broad enough to be analyzed as an ecosystem, and published results show that its execution model can support serious accelerator and distributed workloads. The strength of that evidence is not uniform. The case is most convincing for regular-grid accelerator solvers, element-local high-order methods, and selected distributed applications, where studies report complete solver behavior rather than isolated kernels. Evidence is thinner for general unstructured engineering workflows, robust steady-state solution, industrial multiphysics, and complete inverse or uncertainty-quantification pipelines. This pattern matters because the better documented cases are also those most naturally aligned with compiler specialization and portable kernels. The demonstrated capability therefore applies to specific regions of the CFD design space; breadth, validation history, and operational support still fall well short of established environments.

The more consequential result is the asymmetry between \emph{vertical} and \emph{horizontal} composability. Inside carefully designed projects, numerical-model abstractions can reach downward into specialized kernels and upward into integration, differentiation, optimization, or learning. Between independently designed projects, incompatible state, mesh, boundary, and device semantics remain common. Julia has therefore demonstrated the value of a shared language substrate more clearly than it has demonstrated a shared CFD substrate. In this context, ``single language'' means that the application retains one programmable semantic center in which physical models and numerical operators can be inspected and recombined; meshes, communication routines, or thermochemical models may still be supplied by mature external libraries behind stable interfaces. The architectural benefit depends on avoiding repeated conversion, file-mediated coupling, and duplicate representations of the governing model.

This distinction defines where adoption is most compelling. Rewriting a stable industrial solver solely for language uniformity would discard mature infrastructure without a clear gain, whereas a new or rapidly changing research code that must combine unconventional physics, heterogeneous execution, gradients, and learned components faces growing interface and duplicated-kernel costs. A method-development group may value direct access to equations, custom number types, and differentiation even at modest scale; a new HPC application may benefit when CPU/GPU support and distributed execution are designed around shared abstractions from the outset; and an established production code may use Julia selectively for calibration, surrogate components, or experiment orchestration while preserving its validated core. These are distinct adoption profiles rather than stages on a mandatory migration path. Their risks differ as well: new codes can overgeneralize interfaces before numerical requirements are understood, large applications can depend on the least mature backend primitive, and hybrid deployments can accumulate wrappers that recreate the ownership and derivative problems Julia was meant to reduce. A credible project should identify which boundaries are expected to change, measure the cost of those changes, and avoid treating maximum genericity as an end in itself.

Coding agents alter the labor cost of mixed-language systems but not their semantic boundaries. They can generate wrappers, translate source, and repair tests \citep{Yang2024SWEAgent,Guan2025RepoTransAgent}, yet scientific equivalence still includes conservation, stability, convergence, gradient fidelity, data ownership, and parallel performance. This shifts Julia's argument away from avoiding manual transcription alone: its durable value is the possibility of reducing the number of computational contracts that humans and tools must coordinate. At the same time, agents make explicit interfaces and layered tests more valuable because generated changes are easier to validate when equations, operators, backends, and derivative actions have narrow semantic contracts and independent reference tests. Julia's introspection and generic functions can expose these contracts, but dynamic method extension can also enlarge the behavior that must be searched and understood. AI assistance therefore raises the value of architectures whose scientific invariants are executable and whose performance consequences can be measured after modification.

The main missing evidence is consequently at workflow level. Comparisons with AMReX, Firedrake, and JAX-based CFD establish portability, code generation, and differentiability as shared goals across several software paradigms; Julia's hypothesis is that integrating these capabilities through a general-purpose language reduces total engineering cost. Testing that hypothesis requires case studies that include compilation, implementation and verification effort, backend-specific source, gradient validation, inverse iterations, and reproducible deployment. Such studies can document a sequence of controlled changes---adding an equation, a device backend, a derivative action, or a coupled component---and record source modifications, reused tests, compilation and runtime cost, and numerical verification without reducing incomparable projects to one scalar score. Parallel studies could then compare where specialized C++, a DSL, JAX, or Julia absorbs the change. Current evidence motivates this comparison but does not settle it; carrying it out would turn composability from an impression based on elegant examples into a falsifiable property of evolving scientific software.

\section{Conclusions}
\label{sec:conclusions}

Julia is now a credible implementation environment for several research-facing classes of fluid simulation. The survey evidence includes different discretizations, physical models, accelerator strategies, distributed communication patterns, and differentiated workflows. Its architectural significance lies in how its type system, specialization model, and package interfaces can reduce boundaries between model development, high-performance execution, and inverse or learning algorithms.

That proposition has clear limits. The ecosystem remains smaller and more fragmented than established alternatives; industrial preprocessing and model coverage are incomplete; deployment and cross-vendor tuning require deliberate work; and expressing differentiation or multiphysics coupling in one language does not resolve their algorithmic difficulty. Julia is most persuasive where requirements are evolving and the ability to recombine numerical, hardware, and inference components is itself valuable.

The next phase should evaluate complete computational workflows, including both kernel behavior and the surrounding computation. Time to first solution, verified-gradient cost, effort to add a model or backend, source and tests reused across variants, distributed solver-plus-AD performance, and reproducible deployment would make the productivity--performance claim testable. Julia's broader contribution may be to serve as a demanding laboratory for CFD architectures in which simulation, differentiation, optimization, uncertainty quantification, and learning are designed to interact from the outset.

\section*{Acknowledgments}
The author thanks the developers and contributors of the open-source Julia scientific-computing and CFD projects discussed in this article.

\bibliographystyle{plainnat}
\bibliography{references}

\end{document}